\documentclass[nofootinbib,prd,12pt,superscriptaddress]{revtex4}%
\usepackage{amsmath}
\usepackage{amsfonts}
\usepackage{amssymb}
\usepackage{graphicx}%
\usepackage{amsmath,amssymb}
\usepackage{mathrsfs}
\usepackage{graphicx}
\usepackage{color}
\usepackage{subfigure}
\usepackage{fancyhdr}
\usepackage{multirow}
\usepackage{float}
\usepackage{epsfig}
\usepackage{amsfonts}
\usepackage{bm}
\def\be{\begin{equation}}
\def\ee{\end{equation}}
\def\ba{\begin{eqnarray}}
\def\ea{\end{eqnarray}}

\begin{document}

\title{Observational Constraints on Asymptotic Safety Inflation \\in Gravity’s Rainbow}

\author{Phongpichit Channuie}
\email{phongpichit.ch@mail.wu.ac.th}
\affiliation{School of Science, Walailak University, Nakhon Si Thammarat, 80160, Thailand}
\affiliation{College of Graduate Studies, Walailak University, Nakhon Si Thammarat, 80160, Thailand}

\date{\today}

\begin{abstract}
Using suitable Renormalization Group (RG) based re-summation of quantum corrections to $R^2$ term, a re-summed version of the effective Lagrangian can be obtained \cite{Demmel:2015oqa}. In the context of gravity as an Asymptotically Safe (AS) theory, authors of Refs.\cite{Liu:2018hno,Koshelev:2022olc} proposed a refined Starobinsky model, $L_{\rm AS} = M^{2}_{p} R/2+(\alpha/2)R^{2}/[1+\beta \ln(R/\mu^{2})]$, where $R$ is the Ricci scalar, $\alpha$ and $\beta$ are constants and $\mu$ is an energy scale. In the present work, we embed this underlying effective Lagrangian within the framework of gravity's rainbow. By implementing the COBE normalization and the Planck constraint on the scalar spectrum, we demonstrate that the power spectrum of curvature perturbation relies on $\alpha$ and $\beta$, as well as on a rainbow parameter. Similarly, the scalar spectral index $n_s$ is influenced by $\beta$ and the rainbow parameter, yet remains unaffected by $\alpha$. Additionally, the tensor-to-scalar ratio $r$ solely depends on the rainbow parameter. Remarkably, when requiring $n_s$ to be consistent with the Planck collaboration at $1\sigma$ confidence level, the upper limit on the tensor-to-scalar ratio $r < 0.036$ can be naturally satisfied. This value potentially holds promise for potential measurement by Stage IV CMB ground experiments and is certainly within reach of future dedicated space missions.

\end{abstract}

\maketitle

%%%%%%%%%%%%%%%%%%%%%%%%%%%%%%%%%%%%%%%%

%%%%%%%%%%%%%%%%%%%%%%%
\section{Introduction}
%%%%%%%%%%%%%%%%%%%%%%%
Einstein's general theory of gravity provides a reliable framework for understanding gravity in low energy (IR) descriptions, while it may require refinement in contexts of very high energy (UV) regimes. It is anticipated that at energy scales approaching the Planck scale in various quantum gravity theories, the conventional dispersion relation will undergo modifications. These modifications are particularly relevant in light of research such as that focusing on deformations measured by observations like the Cherenkov array \cite{Cherenkov}. It was first noticed that in Ref.\cite{Magueijo:2002xx}, Magueijo and Smolin proposed the modifications of the dispersion relation, replacing the standard form $\varepsilon^{2}-p^{2}=m^{2}$ with a new expression $\varepsilon^{2}{\tilde f}^{2}(\varepsilon) - p^{2}{\tilde g}^{2}(\varepsilon)=m^{2}$, where ${\tilde f}(\varepsilon)$ and ${\tilde g}(\varepsilon)$ are termed rainbow functions, see also Ref.\cite{Ling:2006az} for inspiration. These functions must adhere to specific conditions, notably approaching unity as energy decreases to an IR limit, represented by ${\tilde f}(\varepsilon/\mu) \rightarrow 1$ and ${\tilde g}(\varepsilon/\mu)\rightarrow 1$, where $\mu$ signifies the energy scale at which quantum gravitational effects become significant.

In the UV limit, the conventional dispersion relation may undergo reformulation to capture a modification of the geometry in that regime. One conceptualization of this notion is to propose that the geometry of spacetime within gravity's rainbow is contingent upon the energy of test particles. Consequently, each test particle, carrying varying energy, perceives a distinct spacetime geometry. This concept gives rise to a family of metrics, known as rainbow metrics, wherein $\varepsilon$ describes the spacetime background instead of a singular metric. Within gravity's rainbow framework, the adjusted metric can be represented as
\ba
g(\varepsilon) = \eta^{\delta\nu}{\tilde e}_{\delta}(\varepsilon)\otimes{\tilde e}_{\nu}(\varepsilon)\,,
\label{act}
\ea
where the energy-dependence of the frame field ${\tilde e}_{\nu}(\varepsilon)$ can be expressed in terms of the energy-independent frame field as ${\tilde e}_{0}(\varepsilon)= e_{0}/{\tilde f}(\varepsilon)$ and ${\tilde e}_{i}(\varepsilon)= e_{i}/{\tilde g}(\varepsilon)$, where $i=1,2,3$. From a cosmological point of view, the standard FLRW spacetime metric describing a homogeneous and isotropic universe is modified using a rainbow metric given by
\begin{eqnarray}
{ds}^2(\varepsilon) =-\frac{dt^2}{\tilde{f}^2(\varepsilon)}+\frac{a^2(t)}{\tilde{g}^2(\varepsilon)} \delta_{ij}{dx}^i{dx}^j\,,\label{modFR}
\end{eqnarray}
where $a(t)$ represents a scale factor. In recent years, the concept of gravity's rainbow has gained significant attention and become a focal point in the literature. Numerous publications have explored various physical aspects of black holes, including \cite{Feng:2016zsj,Hendi:2016hbe,Feng:2017gms,Hendi:2018sbe,Panahiyan:2018fpb,Dehghani:2018qvn,Upadhyay:2018vfu,Dehghani:2018svw,Hendi:2017pld,Hendi:2016njy,Hendi:2015cra,Hendi:2015hja,Ali:2015iba,Ali:2014zea,Ali:2014qra,Ali:2014yea,Ali:2014cpa,EslamPanah:2018ums}. Additionally, the impacts of rainbow functions have been investigated in various cosmological scenarios, such as \cite{Momeni:2017cvl,Deng:2017umx,Khodadi:2016aop,Khodadi:2016bcx,Rudra:2016alu,Ashour:2016cay,Garattini:2012ec,Garattini:2014rwa}. Gravity's rainbow has been also examined within frameworks such as Gauss-Bonnet gravity \cite{Hendi:2016tiy}, massive gravity \cite{Hendi:2017vgo,Heydarzade:2017rpb}, and $f(R)$ gravity \cite{Hendi:2016oxk}. Specifically, it has been applied to analyze the effects of rainbow functions on the Starobinsky model of $f(R)$ gravity \cite{Chatrabhuti:2015mws}. More recently, the deformed Starobinsky model \cite{Codello:2014sua} has also been pursued within the context of gravity's rainbow \cite{Channuie:2019kus}. The study also included the \cite{Waeming:2020rir}.

In this work, we consider an asymptotic safety theory on inflation within the
framework of gravity’s rainbow. The modification of $R^{2}$ inflation with higher curvature is motivated both by considerations in quantum gravity and from a phenomenological point of view. Recent studies, e.g., Refs.\cite{Ferrara:2013rsa,Asaka:2015vza}, have extensively explored extensions of the $R^{2}$ model with higher curvature terms like $R^{n}$ \cite{Asaka:2015vza,Huang:2013hsb,Ivanov:2021chn,Motohashi:2014tra,Bamba:2015uma}. Moreover, the asymptotic safety (AS) approach plays a significant role in shaping this scenario. In particular, the application of renormalization group (RG) techniques for resumming quantum corrections to the $R^{2}$ term has shown promise \cite{Demmel:2015oqa}. In the present work, we take the following $f(R)$ form:
\ba\label{loa}
f_{\rm AS}(R) = R + \frac{\alpha R^{2}}{1+\beta \log\big(\tfrac{R}{\mu^{2}}\big)}\,,
\ea
where $R$ is the Ricci scalar, $\alpha$ and $\beta$ are constants and $\mu$ is an energy scale. Indeed, comparing to Ref.\cite{Demmel:2015oqa}, a parameter $\alpha$ given in Eq.(\ref{loa}) is indeed $\alpha\rightarrow \alpha/M^{2}_{p}$. Note here that when setting $\beta=0$ and $\alpha=1/(6M^{2})$, this model is reduced to the Starobinsky model. According to the findings in Ref.\cite{Liu:2018hno}, which computed inflationary observables for this action, it is revealed that when $b\ll 10^{-3}$, the inflationary predictions closely resemble those of the $R^2$ model. However, for $b\geq 10^{-3}$, there is a slight deviation in predictions, with the tensor-to-scalar ratio potentially reaching as high as $r \sim 10^{-2}$. The validation of these predictions may be possible through future observations, particularly in the detection of B-modes \cite{CMB-S4:2020lpa}. Since asymptotic safety is a relativistic quantum field theory, the standard relativistic dispersion relation is expected to hold in many cases, especially at low energies. However, at the energy scales relevant to inflation, the theory predicts potential modifications to the dispersion relation due to quantum gravitational effects. Consequently, the idea of incorporating rainbow functions into asymptotically safe quantum gravity is physically plausible.

This paper is structured as follows: In Section (\ref{sec1}), we establish a framework for $f(R)$ theory within the paradigm of gravity's rainbow, drawing upon existing reviews \cite{Sotiriou:2008rp,DeFelice:2010aj}. We adopt the $f(R)$ model expressed as $f_{\rm AS}(R) = R/2+(\alpha/2)R^{2}/[1+\beta \ln(R/\mu^{2})]$. In Section (\ref{sec2}), we investigate a snapshot of cosmological linear perturbations arising within the context of gravity's rainbow. Here, we present the spectral index of scalar perturbations and the tensor-to-scalar ratio of the model. Additionally, we juxtapose our predicted outcomes with data from Planck 2018 within this section. Finally, we highlight our findings in the concluding section.

%%%%%%%%%%%%%%%%%%%%%%%
\section{Setup}\label{sec1}
%%%%%%%%%%%%%%%%%%%%%%%
Einstein's theory of gravity, while fundamental, faces significant modern challenges such as dark matter, dark energy, and cosmic inflation. Modifications to general relativity are thus anticipated, especially in the early universe where corrections to Einstein's theory may emerge at high curvature. One straightforward modification involves replacing the Einstein-Hilbert term with a function of the Ricci scalar, giving rise to $f(R)$ theories. Earlier works, including pioneering studies on $f(R)$ and other gravity theories, \cite{Nojiri:2010wj,Nojiri:2017ncd}, have laid the groundwork, see also Ref.\cite{Odintsov:2023weg} for the latest review on modified gravity cosmology covering many modern aspects of modified gravity in early
Universe. Here, we begin our investigation with the standard 4-dimensional action in $f(R)$ gravity, incorporating matter fields \cite{Sotiriou:2008rp,DeFelice:2010aj}.
\ba
S = \frac{1}{2\kappa^2}\int d^4x \sqrt{-g}f(R) + \int d^4x \sqrt{-g}{\cal L}_{M}(g_{\mu\nu},\Psi_{M})\,,
\label{act}
\ea
where we have defined $\kappa^{2}=8\pi G=8\pi/M^{2}_{p}$, $g$ is the determinant of the metric $g_{\mu\nu}$, and the matter field Lagrangian ${\cal L}_{M}$ depends on $g_{\mu\nu}$ and matter fields $\Psi_{M}$. The field equation can be directly obtained by performing a variation of the action (\ref{act}) with respect to $g_{\mu\nu}$ \cite{Sotiriou:2008rp,DeFelice:2010aj}
\ba
F(R)R_{\mu\nu}(g) - \frac{1}{2}f(R)g_{\mu\nu}-\nabla_{\mu}\nabla_{\nu}F(R)+g_{\mu\nu}\Box F(R) = \kappa^{2}T^{(M)}_{\mu\nu}\,,
\label{eom}
\ea
where $F(R)=\partial f(R)/\partial R$ and the operator $\Box$ is defined by $\Box\equiv (1/\sqrt{-g})\partial_{\mu}(\sqrt{-g}g^{\mu\nu}\partial_{\nu})$. Basically, the energy-momentum tensor of the matter fields is given by a definition $T^{(M)}_{\mu\nu}=(-2/\sqrt{-g})\delta(\sqrt{-g}{\cal L}_{M})/\delta g^{\mu\nu}$. Here it satisfies the continuity equation such that $\nabla^{\mu}T^{(M)}_{\mu\nu}=0$. As part of the standard procedure, it's noteworthy that the energy-momentum tensor of matter takes the perfect fluid form: $T^{(M)}_{\mu\nu}={\rm diag}(-\rho,P,P,P)$, where $\rho$ and $P$ denote the energy density and pressure, respectively. Now, we proceed to derive cosmological solutions to the field equations (\ref{eom}). Substituting the modified FLRW metric (\ref{modFR}) into the field equations (\ref{eom}), and assuming the stress-energy tensor is expressed in terms of the perfect fluid form, we obtain:
\begin{eqnarray}
3 \left(FH^2+H\dot{F} \right) -6FH\frac{  \dot{\tilde{g}}}{\tilde{g}} +3 F\frac{ \dot{\tilde{g}}^2}{\tilde{g}^2}+\dot{F}\frac{ \dot{\tilde{f}}}{\tilde{f}}-3 \dot{F}\frac{ \dot{\tilde{g}}}{\tilde{g}}=\frac{F R-f(R)}{2 \tilde{f}^2}+\frac{\kappa ^2 \rho }{\tilde{f}^2}\,,\label{Hdd}
\end{eqnarray}
and
\begin{eqnarray}
&&3 F H^2-3 \dot{F} H+3 F \dot{H}+3 F H\frac{ \dot{\tilde{f}}}{\tilde{f}}-\dot{F}\frac{ \dot{\tilde{f}}}{\tilde{f}}-4 F\frac{ \dot{\tilde{g}}^2}{\tilde{g}^4}+6 F H\frac{ \dot{\tilde{g}}}{\tilde{g}^3}-3 \dot{F}\frac{ \dot{\tilde{g}}}{\tilde{g}^3}+F\frac{ \ddot{\tilde{g}}}{\tilde{g}^3}+F\frac{\dot{\tilde{f}} }{\tilde{f}}\frac{\dot{\tilde{g}}}{\tilde{g}^3}-3 F H^2\frac{1}{\tilde{g}^2}\nonumber\\&&+2 \dot{F} H\frac{1}{\tilde{g}^2}+\frac{\ddot{F}}{\tilde{g}^2}-F \dot{H}\frac{1}{\tilde{g}^2}+6 F\frac{ \dot{\tilde{g}}^2}{\tilde{g}^2}-F H\frac{ \dot{\tilde{f}}}{\tilde{f} }\frac{1}{\tilde{g}^2}+\dot{F}\frac{ \dot{\tilde{f}}}{\tilde{f}}\frac{1}{\tilde{g}^2}-6 F H\frac{ \dot{\tilde{g}}}{\tilde{g}}+3 \dot{F}\frac{ \dot{\tilde{g}}}{\tilde{g}}-3 F\frac{ \ddot{\tilde{g}}}{\tilde{g}}-3 F\frac{\dot{\tilde{f}} }{\tilde{f} }\frac{\dot{\tilde{g}}}{\tilde{g}} \nonumber\\&&-\frac{f(R) \left(\tilde{g}-1\right) \left(\tilde{g}+1\right)}{2 \tilde{f}^2 \tilde{g}^2}=-\frac{\kappa ^2 \left(\rho  \tilde{g}^2+P\right)}{\tilde{f}^2 \tilde{g}^2}\,,\label{ijcom}
\end{eqnarray}
where we have defined a first and second derivative with respect to time with ${\dot a}$ and ${\ddot a}$, respectively. For simplicity, in our analysis below we chose $\tilde{g}=1$ and only considered the spatially flat universe. From the equation (\ref{loa}), we obtain
\begin{eqnarray}
F_{\rm AS}(R)=f_{\rm AS}'(R)&\equiv& \frac{\partial f_{\rm AS}(R)}{\partial R}= 1+\frac{2 \alpha  R}{\beta  \log \left(\frac{R}{\mu^2}\right)+1}-\frac{\alpha  \beta  R}{\left(\beta  \log \left(\frac{R}{\mu^2}\right)+1\right)^2}\,,\\
f_{\rm AS}''(R)\equiv \frac{\partial^{2} f_{\rm AS}(R)}{\partial R^{2}}&=&\frac{2 \alpha  \beta ^2}{\left(\beta  \log \left(\frac{R}{\mu ^2}\right)+1\right)^3}-\frac{3 \alpha  \beta }{\left(\beta  \log \left(\frac{R}{\mu ^2}\right)+1\right)^2}+\frac{2 \alpha }{\beta  \log \left(\frac{R}{\mu ^2}\right)+1}\,,
\end{eqnarray}
The function $f_{\rm AS}(R)$ obeys the quantum stability condition $f''_{\rm AS}(R)>0$ for $\alpha >0$ and $\beta >0$. This ensures the stability of the solution at high curvature. Additionally, the condition of classical stability leads to
\begin{eqnarray}
f_{\rm AS}'(R) = 1+\frac{2 \alpha  R}{\beta  \log \left(\frac{R}{\mu^2}\right)+1}-\frac{\alpha  \beta  R}{\left(\beta  \log \left(\frac{R}{\mu^2}\right)+1\right)^2}>0\,.
\end{eqnarray}
Note that the functions 
${\tilde{f}(\varepsilon)}$ and ${\tilde{g}(\varepsilon)}$ modify the spacetime metric in an energy-dependent manner. These functions are not uniquely determined by the theory and are chosen based on theoretical considerations and consistency with observations. Importantly, the chosen forms for 
${\tilde{f}(\varepsilon)}$ and ${\tilde{g}(\varepsilon)}$ should reduce to $1$ at low energies ($\varepsilon\ll 1$) to ensure that classical general relativity is recovered in the low-energy limit. In this work we assume that $\tilde{f} =1+(H/\mu)^{\lambda}$ where $\lambda$ is called "a rainbow parameter" and $\mu$ is the energy scale that quantum effects of gravity become important. In an inflationary regime, where $H\gg \mu$, the function $\tilde{f}$ can be approximated as $\tilde{f} \approx (H/\mu)^{\lambda}$. From Eq.(\ref{Hdd}), we find for this model
\begin{eqnarray}
&&-\frac{3}{H \mu  \left(\beta  \log \left(\frac{12 H^2 \left(\frac{H}{\mu }\right)^{2 \lambda }}{\mu ^2}\right)+1\right)^3} \Bigg(-24 \alpha  H^3 (\lambda +1) \mu  \left(\frac{H}{\mu }\right)^{2 \lambda } {\dot H}(t) \Bigg(\beta  (2 \beta -3)\nonumber\\&&+\beta  \log \left(\frac{12 H^2 \left(\frac{H}{\mu }\right)^{2 \lambda }}{\mu ^2}\right) \left(-3 \beta +2 \beta  \log \Bigg(\frac{12 H^2 \left(\frac{H}{\mu }\right)^{2 \lambda }}{\mu ^2}\right)+4\Bigg)+2\Bigg)\nonumber\\&&-H^3 \mu  \left(\beta  \log \left(\frac{12 H^2 \left(\frac{H}{\mu }\right)^{2 \lambda }}{\mu ^2}\right)+1\right) \Bigg(\left(\beta  \log \left(\frac{12 H^2 \left(\frac{H}{\mu }\right)^{2 \lambda }}{\mu ^2}\right)+1\right)^2\nonumber\\&&+6 \alpha  \left(\frac{H}{\mu }\right)^{2 \lambda } \left(2 H^2+(\lambda +1) {\dot H}(t)\right) \left(-\beta +2 \beta  \log \left(\frac{12 H^2 \left(\frac{H}{\mu }\right)^{2 \lambda }}{\mu ^2}\right)+2\right)\Bigg)\Bigg)\nonumber\\&=&\frac{18 \alpha  \left(\frac{H}{\mu }\right)^{2 \lambda } \left(4 H^4+4 H^2 (\lambda +1) {\dot H}\right) \left(-\beta +\beta  \log \left(\frac{12 H^2 \left(\frac{H}{\mu }\right)^{2 \lambda }}{\mu ^2}\right)+1\right)}{\left(\beta  \log \left(\frac{12 H^2 \left(\frac{H}{\mu }\right)^{2 \lambda }}{\mu ^2}\right)+1\right)^2}\,,\label{2sft}
\end{eqnarray}
and from (\ref{ijcom})
\begin{eqnarray}
&&\frac{1}{\left(\beta  \log \left(\frac{R}{\mu ^2}\right)+1\right)^4}\Bigg(\left(\beta  \log \left(\frac{R}{\mu ^2}\right)+1\right) \Bigg(2 H' \left(\beta  \log \left(\frac{R}{\mu ^2}\right)+1\right)^3\nonumber\\&&-\alpha  \left(H R'-R''\right) \left(\beta  (2 \beta -3)+\beta  \log \left(\frac{R}{\mu ^2}\right) \left(-3 \beta +2 \beta  \log \left(\frac{R}{\mu ^2}\right)+4\right)+2\right)\Bigg)\nonumber\\&&+2 \alpha  R H' \left(-\beta +2 \beta  \log \left(\frac{R}{\mu ^2}\right)+2\right) \left(\beta  \log \left(\frac{R}{\mu ^2}\right)+1\right)^2\nonumber\\&&-\frac{2 \alpha  \beta  \left(R'\right)^2 \left(3 (\beta -1) \beta +\beta  \log \left(\frac{R}{\mu ^2}\right) \left(-3 \beta +\beta  \log \left(\frac{R(t)}{\mu ^2}\right)+2\right)+1\right)}{R}\Bigg)\nonumber\\&&+2 H \Bigg(\frac{\alpha  R \left(-\beta +2 \beta  \log \left(\frac{R}{\mu ^2}\right)+2\right)}{\left(\beta  \log \left(\frac{R}{\mu ^2}\right)+1\right)^2}+1\Bigg)\frac{\dot{\tilde f}}{{\tilde f}}=0\,. \label{22sft}
\end{eqnarray}
Here we are only interested in an inflationary solution. Therefore we invoke the slow-roll approximations. Hence the terms containing $\ddot{H}$ and higher power in $\dot{H}$ can be neglected in this particular regime. It is rather straightforward to show that the Eq.(\ref{2sft}) is reduced to
\begin{eqnarray}
&\dot{H}&\simeq \frac{\left(\frac{H}{\mu }\right)^{-2 \lambda } \left(-12 \alpha  \beta  H^2 \left(\frac{H}{\mu }\right)^{2 \lambda } \left(\Phi(H,\mu)+1\right)-\left(\Phi(H,\mu)+1\right)^3\right)}{6 \alpha  (\lambda +1) \left(\beta  (8 \beta -9)+3 \Phi(H,\mu) \left(-3 \beta +2 \Phi(H,\mu)+4\right)+6\right)},\label{HdHumod2}
\end{eqnarray}
where we have defined a new function $\Phi(H,\mu)$:
\begin{eqnarray}
\Phi(H,\mu)\equiv \beta  \log \Bigg(\frac{12 H^2 \Big(\frac{H}{\mu }\Big)^{2 \lambda }}{\mu ^2}\Bigg)\,.
\end{eqnarray}
Note that when setting $\beta=0$ and $\alpha=1/(6M^{2})$ the result converts to that of Ref.\cite{Chatrabhuti:2015mws}:
\ba
\dot{H}\to -\frac{M^2 \left(\frac{H}{M}\right)^{-2 \lambda }}{6 (\lambda +1)}\,.
\ea
During inflation we can assume $H\simeq {\rm constant.}$, and then in this situation we obtain from Eq.(\ref{2sft})
\ba
H &\simeq& H_i-\frac{\left(\frac{H_i}{\mu }\right)^{-2 \lambda }\left(\left(1+\Phi \left(H_i,\mu \right)\right){}^3+12\alpha \beta \left(1+\Phi \left(H_i,\mu \right)\right)H_i^2\left(\frac{H_i}{\mu }\right){}^{2 \lambda }\right)}{6\alpha \lambda +1\left(6+\beta  (8 \beta -9)+3 \Phi \left(H_i,\mu \right) \left(4-3 \beta +2 \Phi \left(H_i,\mu \right)\right)\right)}(t-t_i)\,,\label{Htkmod2}
\ea
and 
\ba
a \simeq a_i \exp \Bigg\{H_i(t-t_i)-\Gamma(\Phi_{i})\frac{(t-t_i)^{2}}{2}\Bigg\},
\ea
where $H_i$ and $a_i$ are respectively the Hubble parameter and the scale factor at the onset of inflation ($t=t_i$) and we have defined a new function $\Gamma(\Phi_{i})$:
\ba
\Gamma(\Phi_{i}) \equiv \frac{\left(\frac{H_i}{\mu }\right)^{-2 \lambda }\left(\left(1+\Phi \left(H_i,\mu \right)\right){}^3+12\alpha \beta \left(1+\Phi \left(H_i,\mu \right)\right)H_i^2\left(\frac{H_i}{\mu }\right){}^{2 \lambda }\right)}{6\alpha \lambda +1\left(6+\beta  (8 \beta -9)+3 \Phi \left(H_i,\mu \right) \left(4-3 \beta +2 \Phi \left(H_i,\mu \right)\right)\right)}.
\ea
The slow-roll parameter $\varepsilon_1$ is defined by $\varepsilon_1 \equiv -\dot{H}/H^2$ which in this case can be estimated as
\begin{eqnarray}
\varepsilon_{1}\equiv -\frac{\dot{H}}{H^{2}}\simeq \frac{\left(\frac{H}{\mu }\right)^{-2 \lambda }\left(12 H^2 \alpha \beta \left(\frac{H}{\mu }\right)^{2 \lambda }(1+\Phi (H,\mu))+(1+\Phi (H,\mu ))^3\right)}{6 H^2 \alpha \lambda +1(6+\beta  (8 \beta -9)+3 \Phi(H,\mu)(4-3 \beta +2\Phi(H,\mu )))}.\label{epumod2}
\end{eqnarray}
We can check that $\varepsilon_{1}$ is less than unity during inflation ($H \gg \mu$) and we find when setting $\beta=0,\,\alpha=1/(6M^{2})$ that the above expression reduces to $\varepsilon_1 \simeq\frac{H^{-2 (\lambda +1)} M^{2 \lambda +2}}{6 (\lambda +1)}$. One can simply determine the time when inflation ends ($t=t_f$) by solving $\varepsilon(t_f) \simeq 1$ to obtain
\ba
t_f &\simeq& t_i + \frac{6 \alpha \lambda +\left(6+\beta  (8 \beta -9)+3\Phi \left(H_i,\mu \right)\left(4-3 \beta +2 \Phi \left(H_i,\mu \right)\right)\right) H_i \left(\frac{H_i}{\mu }\right){}^{2 \lambda}}{-\left(1+\Phi \left(H_i,\mu \right)\right)^3-12 \alpha \beta \left(1+\Phi \left(H_i,\mu \right)\right)H_i^2 \left(\frac{H_i}{\mu }\right)^{2 \lambda}}\, . \label{tfmod1}
\ea
The number of e-foldings from $t_i$ to $t_f$ is then given by
\ba
N &\equiv& \int^{t_f}_{t_i} Hdt \nonumber\\&\simeq& H_i(t-t_i)-\frac{\left(\frac{H_i}{\mu }\right)^{-2 \lambda }\left(\left(1+\Phi \left(H_i,\mu \right)\right){}^3+12\alpha \beta \left(1+\Phi \left(H_i,\mu \right)\right)H_i^2\left(\frac{H_i}{\mu }\right){}^{2 \lambda }\right)}{6\alpha \lambda +1\left(6+\beta  (8 \beta -9)+3 \Phi \left(H_i,\mu \right) \left(4-3 \beta +2 \Phi \left(H_i,\mu \right)\right)\right)}\frac{(t-t_i)^{2}}{2}\nonumber\\&\simeq& \frac{1}{2\varepsilon_{1}(t_{i})}\, .
\ea
Note that when $\alpha=1/(6M^{2})$  and $\beta=0$, the result is the same as that of the Starobinsky model. In the following section, we examine the spectra of perturbations within the framework of gravity's rainbow theory. We then confront the results predicted by our models with Planck 2018 data.

%%%%%%%%%%%%%%%%%%%%%%%
\section{Confrontation with Observation}\label{sec2}
\begin{figure}[ht!]
    \centering
\includegraphics[width=5in,height=5in,keepaspectratio=true]{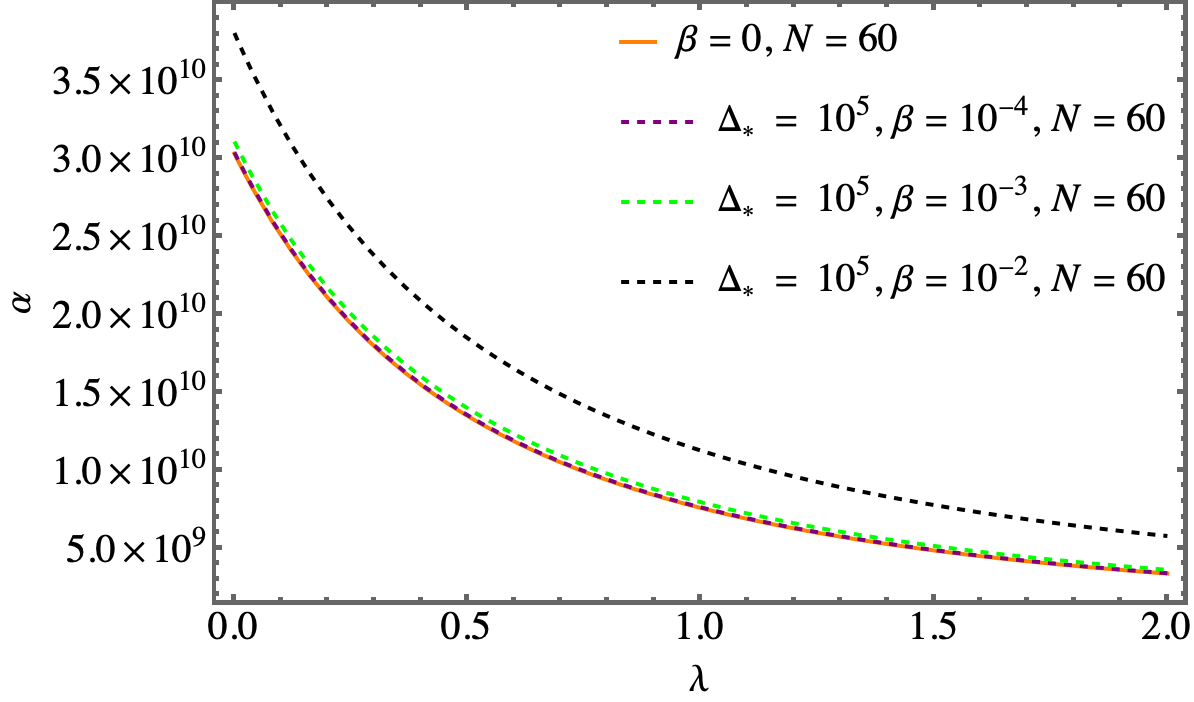}
\includegraphics[width=5in,height=5in,keepaspectratio=true]{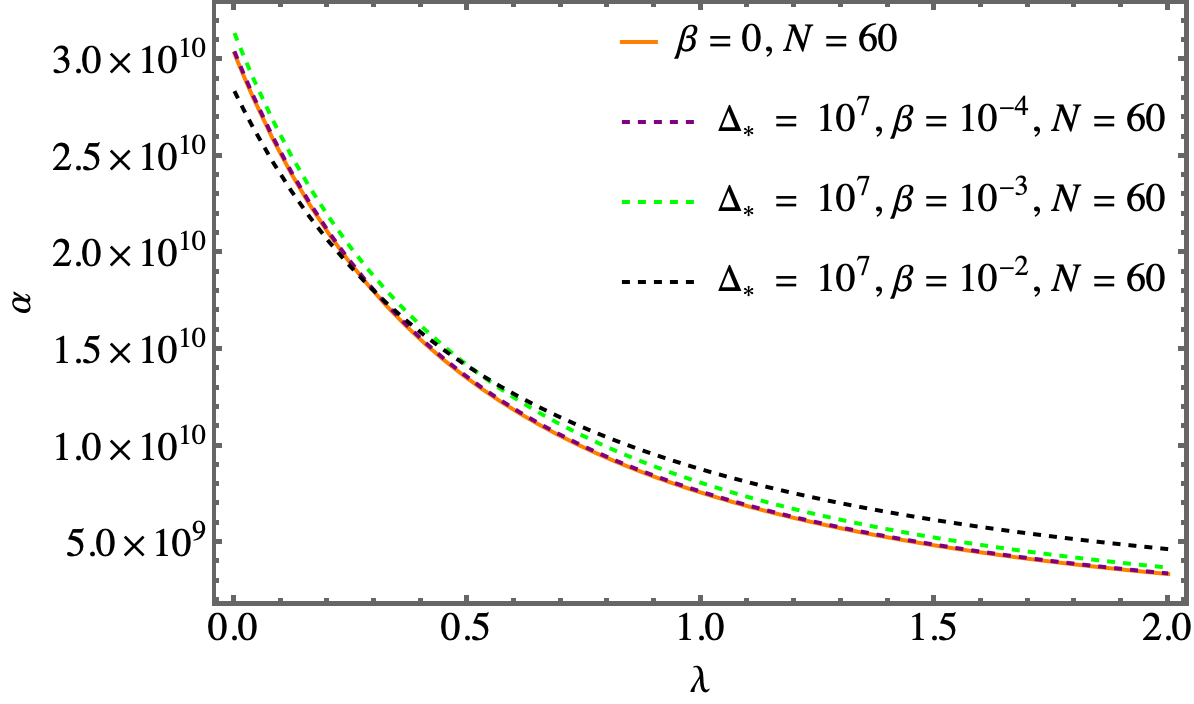}
    \caption{We present the behavior of $\alpha$ versus $\lambda$, evaluated at the Hubble radius crossing $(k = aH)$ in Eq. (\ref{alpha}), for $\beta$ values of $[10^{-4}, 10^{-3}, 10^{-2}]$ while keeping $N = 60$ fixed. The top panel shows results for $\Delta_{*} = 10^5$, and the bottom panel shows results for $\Delta_{*} = 10^7$. For reference, we also include a plot for $\beta = 0$ with $N = 60$.}
    \label{figal}
\end{figure}

%%%%%%%%%%%%%%%%%%%%%%%
It is worth mentioning that while quadratic gravity is highly successful in describing inflation, combining it with gravity’s rainbow could (I) examine the effects of modified dispersion relations on inflationary perturbations and (II) generate new, testable predictions for cosmological observations. In this section, we closely follow Refs.\cite{Channuie:2019kus,Waeming:2020rir} for a cosmological linear perturbation in the context of the gravity’s rainbow generated during inflation. Here the reader can find a detailed derivation of the spectral index of curvature perturbation and the tensor-to-scalar ratio. In this section, we will briefly review cosmological linear perturbation within the framework of gravity's rainbow during inflation, as proposed by Ref.\cite{Chatrabhuti:2015mws}. We start with scalar perturbation (since scalar and tensor perturbations evolve independently at the linear level) using the following perturbed flat FRW metric that incorporates the rainbow effect:
\ba
ds^2 = -\frac{1+2\Phi}{\tilde{f}^2(t)}dt^2 + a^2(t)(1-2\Psi)d\vec{x}^2 , , \label{pertFRW}
\ea
where $\tilde{f}(t)$ is the rainbow function. Note that this perturbed metric is expressed in the Newtonian gauge. We introduce a new variable $A \equiv 3(H\Phi+\dot{\Psi})$. With the metric (\ref{pertFRW}) and Eq.(\ref{eom}), we derive the following system of equations \cite{Chatrabhuti:2015mws}:
\ba
-\frac{\nabla^2\Psi}{a^2}+\tilde{f}^2HA &=& -\frac{1}{2F}\left[3\tilde{f}^2\left(H^2 + \dot{H} + \frac{\dot{\tilde{f}}}{\tilde{f}}\right)\delta F + \frac{\nabla^2\delta F}{a^2}-3\tilde{f}^2H\delta\dot{F} \right. \nonumber \\ 
 &+& \left. 3\tilde{f}^2H\dot{F}\Phi + \tilde{f}^2\dot{F}A+\kappa^2\delta\rho_M\right] \, , \label{eomA1} \\
H\Phi+\dot{\Psi}&=&-\frac{1}{2F}(H\delta F+\dot{F}\Phi-\delta\dot{F}) \, ,  \label{eomA2} 
\ea
and 
\ba
\dot{A} + \left(2H+\frac{\dot{\tilde{f}}}{\tilde{f}}\right)A+3\dot{H}\Phi + \frac{\nabla^2\Phi}{a^2\tilde{f}^2}+\frac{3H\Phi\dot{\tilde{f}}}{\tilde{f}} 
= \frac{1}{2F}\left[3\delta\ddot{F}+3\left(H+\frac{\dot{\tilde{f}}}{\tilde{f}}\right)\delta\dot{F} \right. \nonumber \\ -\left. 6H^2\delta F -\frac{\nabla^2\delta F}{a^2\tilde{f}^2} - 3\dot{F}\dot{\Phi}-\dot{F}A - 3\left(H+\frac{\dot{\tilde{f}}}{\tilde{f}}\right)\dot{F}\Phi-6\ddot{F}\Phi+\frac{\kappa^2}{\tilde{f}^2}(3\delta P_M+\delta\rho_M) \right] \, .
\label{eomA3}
\ea
Note that the above equations can be used to describe the evolution of cosmological scalar perturbations. In the following, we will solve these equations within the inflationary framework. We first examine scalar perturbations generated during inflation and assume a perfect fluid is not present, i.e., $\delta\rho_M = 0$ and $\delta P_M = 0$. We select the gauge condition $\delta F = 0$, so that $\mathcal{R} = \psi = -\Psi$. It is important to note that the spatial curvature $^{(3)}\mathcal{R}$ on the constant-time hypersurface is related to $\psi$ by the relation $^{(3)}\mathcal{R} = - 4\nabla^2\psi/a^2$. Using $\delta F = 0$, we derive from Eq.(\ref{eomA2}) that:
\begin{eqnarray}
\Phi = \frac{\dot{\mathcal{R}}}{H + \dot{F}/2F} \ , \label{B8}
\end{eqnarray}
and from Eq.(\ref{eomA1}), we find:
\begin{eqnarray}
A = -\frac{1}{H + \dot{F}/2F}\left[\frac{\nabla^2\mathcal{R}}{a^2\tilde{f}^2} + \frac{3H\dot{F}\dot{\mathcal{R}}}{2F(H + \dot{F}/2F)}\right] \ . \label{B9}
\end{eqnarray}
Using the background equation (\ref{ijcom}), we obtain from Eq.(\ref{eomA3}):
\begin{align}
\dot{A} + \left(2H + \frac{\dot{F}}{2F}\right)A + \frac{\dot{\tilde{f}}A}{\tilde{f}} + \frac{3\dot{F}\dot{\Phi}}{2F} + \left[\frac{3\ddot{F} + 6H\dot{F}}{2F} + \frac{\nabla^2}{a^2\tilde{f}^2}\right]\Phi + \frac{3\dot{F}}{2F}\frac{\Phi\dot{\tilde{f}}}{\tilde{f}} = 0. \label{B10}
\end{align}
Substituting Eq.(\ref{B8}) and (\ref{B9}) into Eq.(\ref{B10}), we find that in Fourier space, the curvature perturbation satisfies the following equation:
\begin{eqnarray}
\ddot{\mathcal{R}} + \frac{1}{a^3Q_s}\frac{d}{dt}(a^3Q_s)\dot{\mathcal{R}} + \frac{\dot{\tilde{f}}}{\tilde{f}}\dot{\mathcal{R}} + \frac{k^2}{a^2\tilde{f}^2} \mathcal{R}= 0 \ , \label{B13}
\end{eqnarray}
where $k$ is a comoving wave number and $Q_s$ is defined by	
\begin{eqnarray}
 Q_s \equiv \frac{3\dot{F}^2}{2\kappa^2F(H+\dot{F}/2F)^2} \ . \label{B12}
\end{eqnarray}
Introducing new variables $z_s = a\sqrt{Q_s}$ and $u = z_s\mathcal{R}$, Eq.(\ref{B13}) can be simplified to:
\begin{eqnarray}
u'' + \left(k^2-\frac{z_s''}{z_s}\right)u = 0 \ , \label{B14}
\end{eqnarray}
where a prime denotes a derivative with respect to the new time coordinate $\eta = \int (a\tilde{f})^{-1} dt$. To determine the spectrum of curvature perturbations, we define slow-roll parameters as:
\begin{eqnarray}
 \epsilon_1 \equiv -\frac{\dot{H}}{H^2},  \ \ \epsilon_2 \equiv \frac{\dot{F}}{2HF}, \ \ \epsilon_3 \equiv \frac{\dot{E}}{2HE}\ ,
\end{eqnarray}
where $E \equiv 3\dot{F}^2/2\kappa^2$. Consequently, $Q_s$ can be rewritten as:
\begin{eqnarray}
 Q_s = \frac{E}{FH^2(1+\epsilon_2)^2} \ . \label{B16}
\end{eqnarray}
Here, the parameters $\epsilon_i$ are assumed to be nearly constant during inflation, and $\tilde{f} \simeq (H/M)^{\lambda}$. These assumptions allow us to calculate 
$\eta$ as $\eta = -1/[(1-(1+\lambda)\epsilon_1)\tilde{f}aH]$. If $\dot{\epsilon_i}\simeq 0$, a term $z_s''/z_s$ satisfies:
\begin{eqnarray}
\frac{z_s''}{z_s} = \frac{\nu^2_{\mathcal{R}} - 1/4}{\eta^2} \ , \label{B18}
\end{eqnarray}
with:
\begin{eqnarray}
\nu_{\mathcal{R}}^2 = \frac{1}{4} + \frac{(1+\epsilon_1 - \epsilon_2+\epsilon_3)(2-\lambda\epsilon_1 -\epsilon_2+\epsilon_3)}{(1-(\lambda+1)\epsilon_1)^2} \ . \label{B19}
\end{eqnarray}
Therefore we find the solution of Eq.(\ref{B14}) written in terms of a linear combination of Hankel functions
\begin{eqnarray}
 u = \frac{\sqrt{\pi|\eta|}}{2}\textmd{e}^{i(1+2\nu_{\mathcal{R}})\pi/4}\left[c_1\textmd{H}_{\nu_{\mathcal{R}}}^{(1)}(k|\eta|)+c_2\textmd{H}_{\nu_{\mathcal{R}}}^{(2)}(k|\eta|)\right] \ , \label{B20}
\end{eqnarray}
where $c_1$, $c_2$ are integration constants and $\textmd{H}_{\nu_{\mathcal{R}}}^{(1)}(k|\eta|)$, $\textmd{H}_{\nu_{\mathcal{R}}}^{(2)}(k|\eta|)$ are the Hankel functions of the first kind and the second kind respectively.
In the asymptotic past $k\eta \rightarrow -\infty$, we find from Eq.(\ref{B20}) $u \rightarrow \textmd{e}^{-ik\eta}/\sqrt{2k}$. This implies $c_1=1$ and $c_2=0$ giving the following solutions
\begin{eqnarray}
 u = \frac{\sqrt{\pi|\eta|}}{2}\textmd{e}^{i(1+2\nu_{\mathcal{R}})\pi/4}\textmd{H}_{\nu_{\mathcal{R}}}^{(1)}(k|\eta|) \ . \label{B21}
\end{eqnarray}
By defining the power spectrum of curvature perturbations
\begin{eqnarray}
 \mathcal{P}_{\mathcal{R}} \equiv \frac{4\pi k^3}{(2\pi)^3}|\mathcal{R}|^2 \ , \label{B22}
\end{eqnarray}
and using Eq.(\ref{B21}) and $u = z_s\mathcal{R}$, we obtain
\begin{eqnarray}
 \mathcal{P}_{\mathcal{R}} = \frac{1}{Q_s}\left[(1-(1+\lambda)\epsilon_1)\frac{\Gamma(\nu_{\mathcal{R}})H}{2\pi\Gamma(3/2)}\left(\frac{H}{M}\right)^\lambda\right]^2\left(\frac{k|\eta|}{2}\right)^{3-2\nu_{\mathcal{R}}} \ , \label{B23}
\end{eqnarray}
where we have used $\textmd{H}_{\nu_{\mathcal{R}}}^{(1)}(k|\eta|) \rightarrow -(i/\pi)\Gamma(\nu_{\mathcal{R}})(k|\eta|/2)^{-\nu_{\mathcal{R}}}$ for $k|\eta| \rightarrow 0$. Since $\mathcal{R}$ is frozen after the Hubble radius crossing, $P_{\mathcal{R}}$ should be evaluated at $k=aH$. Now we define the spectral index $n_{\mathcal{R}}$  as
\begin{eqnarray}
 n_{s} - 1 = \left.\frac{d\textmd{ln}\mathcal{P}_{\mathcal{R}}}{d\textmd{ln}k}\right|_{k=aH} = 3 - 2\nu_{\mathcal{R}} \ . \label{B25}
\end{eqnarray}
The spectral index can be written in terms of the slow-roll parameters as
\begin{eqnarray}
 n_{s} - 1 \simeq -2(\lambda+2)\epsilon_1+2\epsilon_2-2\epsilon_3 \ , \label{B27}
\end{eqnarray}
where during the inflationary epoch, we have assumed that $|\epsilon_i | \ll 1$. Notice that the spectrum is nearly scale-invariant when $|\epsilon_i|$ are much smaller than unity, i.e. $n_{\mathcal{R}} \simeq 1$.   Subsequently, the power spectrum of curvature perturbation takes the form
\begin{eqnarray}
 \mathcal{P}_{\mathcal{R}} \approx \frac{1}{Q_s}\left(\frac{H}{2\pi}\right)^2\left(\frac{H}{M}\right)^{2\lambda} \ . \label{B28}
\end{eqnarray}
Note that we obtain the standard result when setting $\lambda=0$ \cite{DeFelice:2010aj}. We next consider the tensor perturbation. In general $h_{ij}$ can be generally written as
\begin{eqnarray}
 h_{ij} = h_{+}e^+_{ij} + h_{\times}e^\times_{ij} \ , \label{C2}
\end{eqnarray}
where $e^+_{ij}$ and $e^\times_{ij}$ are the polarization tensors corresponding to the two polarization states of $h_{ij}$. Let $\vec{k}$ be in the direction along the z-axis, then the non-vanishing components of polarization tensors are $e^+_{xx} = -e^+_{yy} = 1$ and $e^\times_{xy} = e^\times_{yx} = 1$. Without taking into account the scalar and vector perturbation, the perturbed FLRW metric can be written as
\begin{align}
 ds^2 = -\frac{dt^2}{\tilde{f}(\varepsilon)^2} + a^2(t)h_{\times}dxdy + a^2(t)\left[(1+h_{+})dx^2+(1-h_{+})dy^2+dz^2\right] .\label{C1}
\end{align}
Using Eq.(\ref{eom}), we can show that the Fourier components $h_\chi$  satisfy the following equation
\begin{eqnarray}
  \ddot{h}_\chi + \frac{(a^3F)^\cdot}{a^3F}\dot{h}_\chi + \frac{\dot{\tilde{f}}}{\tilde{f}}\dot{h}_\chi + \frac{k^2}{a^2\tilde{f}^2}h_\chi = 0 \ , \label{C5}
\end{eqnarray}
where $\chi$ denotes polarizations $+$ and $\times$. Following a similar procedure to  the case of curvature perturbation, let us introduce the new variables $z_t = a\sqrt{F}$ and $u_\chi = z_t h_\chi /\sqrt{2 \kappa^{2}}$. Therefore Eq. (\ref{C5}) can be written as
\begin{eqnarray}
 u''_\chi + \left(k^2-\frac{z_t''}{z_t}\right)u_\chi = 0 \ . \label{C6}
\end{eqnarray}
Notice that for a massless scalar field $u_\chi$ has dimension of mass. By choosing $\dot{\epsilon}_i=0$, we obtain
\begin{eqnarray}
 \frac{z_t''}{z_t} = \frac{\nu^2_t-1/4}{\eta^2} \ , \label{C7}
\end{eqnarray}
where
\begin{eqnarray}
 \nu^2_t = \frac{1}{4} + \frac{(1+\epsilon_2)(2-(1+\lambda)\epsilon_1+\epsilon_2)}{(1-(1+\lambda)\epsilon_1)^2} \ . \label{C8}
\end{eqnarray}
Similarly the solution to Eq.(\ref{C6}) can be also expressed in terms of a linear combination of Hankel functions. Taking into account polarization
states, the power spectrum of tensor perturbations $P_T$ after the Hubble radius crossing reads  
\begin{align}
 \mathcal{P}_T &= 4\times\frac{2\kappa^{2}}{a^2F}\frac{4\pi k^3}{(2\pi)^3}|u_\chi|^2  \nonumber\\&= \frac{16}{\pi}\left(\frac{H}{M_{P}}\right)^2\frac{1}{F}\left[(1-(1+\lambda)\epsilon_1)\frac{\Gamma(\nu_t)}{\Gamma(3/2)}\left(\frac{H}{M}\right)^\lambda\right]^2\left(\frac{k|\eta|}{2} \right)^{3-2\nu_t},  
 \label{C12}
\end{align}
where we have used $\tilde{f} \simeq (H/M)^{\lambda}$. Therefore $\nu_t$ can be estimated by assuming that the slow-roll parameters are very small during inflation as
\begin{eqnarray}
 \nu_t \simeq \frac{3}{2} + (1+\lambda)\epsilon_1 + \epsilon_2 \ . \label{C13}
\end{eqnarray}
In addition, the spectral index of tensor perturbations is determined via
\begin{eqnarray}
 n_T = \left.\frac{d \textmd{ln}\mathcal{P}_T}{d\textmd{ln}k}\right|_{k=aH} = 3-2\nu_t \simeq -2(1+\lambda)\epsilon_1 - 2\epsilon_2 \ . \label{C16}
\end{eqnarray}
The power spectrum $\mathcal{P}_T$ can also be rewritten as
\begin{eqnarray}
 \mathcal{P}_T \simeq \frac{16}{\pi}\left(\frac{H}{M_{P}}\right)^2\frac{1}{F}\left(\frac{H}{M}\right)^{2\lambda} \ . \label{C17}
\end{eqnarray}
Also, the tensor-to-scalar ratio $r$ can be determined by invoking the following definition:
\begin{eqnarray}
 r \equiv \frac{\mathcal{P}_T}{\mathcal{P}_{\cal R}}  \simeq 48\epsilon_2^2 \,. \label{a43}
\end{eqnarray}
In the next section, we consider the spectra of perturbations based on various $f(R)$ models in gravity’s rainbow theory and confront the results predicted by our models with Planck 2018 data. Following Refs.\cite{Channuie:2019kus,Waeming:2020rir}, a relation between $\varepsilon_{1}$ and $\varepsilon_{1}$ can be verified:
\begin{eqnarray}
    \epsilon_2 \simeq -(1+\lambda)\epsilon_1 . \label{ep1ep2}
\end{eqnarray}
We can verify another relation among slow-roll parameters by considering the definition of $\epsilon_3$
\begin{eqnarray}
    \epsilon_3 \equiv \frac{\dot{E}}{2HE}=\frac{\ddot{F}}{H \dot{F}}.
\end{eqnarray}
In order to verify the relations among slow-roll parameters, we will focus on some different forms of $f(R)$ given below. We can show that
\ba
Q_s\simeq \frac{3 M_{p}^2}{4\pi}\varepsilon _2^2 F_{\rm AS}(R)
\ea
We consider Eq.(\ref{B28}) and then the power spectrum of curvature perturbation reads
\begin{eqnarray}
    \mathcal{P}_{\cal R} \approx \frac{1}{Q_s}\left(\frac{H}{2\pi}\right)^2\left(\frac{H}{M}\right)^{2\lambda} = \frac{1}{72 \pi  \alpha M_{p}^2}\Big(1+\beta  \log \Big(12 \Big(\frac{H}{\mu }\Big)^{2 \lambda+2 }\Big)\Big)\frac{1}{(1+\lambda)^{2}\epsilon^{2}_{1}}\,.\label{PRmod1}
\end{eqnarray}
In this model, therefore, $\epsilon_3$ reads
\begin{eqnarray}
    \epsilon_3\simeq \epsilon _1 \Bigg(\frac{4 \beta  (\lambda +1)}{2 \beta  \log \left(12 \Delta ^{2 \lambda +2}\right)-3 \beta +2}-2 \lambda -1\Bigg)\,,\label{ep3mod1}
\end{eqnarray}
where we have assumed the slow-roll approximations so that the terms containing $\ddot{H}$ and a higher power of $\beta$ can be ignored. Notice that the approximated result is independent of $\alpha$. Having used Eq.(\ref{ep1ep2}) and Eq.(\ref{ep3mod1}), hence, we have
\begin{eqnarray}
 n_{s} - 1 &\simeq& -2(\lambda+2)\epsilon_1+2\epsilon_2-2\epsilon_3 \ ,\nonumber\\
&\simeq&\epsilon_1\Bigg(-4-\frac{8 \beta  (\lambda +1)}{2 \beta  \log \left(12 \Delta ^{2 \lambda +2}\right)-3 \beta +2}\Bigg)\ ,
\end{eqnarray}
where we have defined a new parameter $\Delta \equiv H/\mu$. For simplicity, let us suppose that during inflation the expansion is de Sitter (exponential) with a constant Hubble parameter. In terms of the number of efoldings, $\mathcal{P}_{\cal R},\,n_{s}$ and $r$ read
\begin{eqnarray}
    \mathcal{P}_{\cal R} &\approx& \frac{N^2}{18 \pi  \alpha  (\lambda +1)^2 M_{p}^2}+\frac{N^2 \log \left(12 \Delta ^{2 \lambda +2}\right)}{18 \pi  \alpha  (\lambda +1)^2 M_{p}^2}\beta ,\label{PR33m1}\\
    n_{s}-1&\approx&-\frac{2}{N}-\frac{2   (\lambda +1)}{N}\beta+\frac{(\lambda +1) \left(2 \log \left(12 \Delta ^{2 \lambda +2}\right)-3\right)}{N}\beta^2, \label{finalnsm1} \\
    r&\approx& \frac{12 (\lambda +1)^2}{N^2}. \label{finalrm2}
\end{eqnarray}
We find that the above parameters reduce to those of the Starobinsky model when $\lambda=0,\,\alpha=1/6M^{2}$. We then take the latest Planck2018 data \cite{Planck:2018jri,Planck:2018vyg} and the latest BICEP/Keck data \cite{BICEP:2021xfz} for well-defined parameters, the scalar spectral index $n_s$ and the tensor-to-scalar ratio $r$, as follows:
\begin{eqnarray}
    &&{\rm Planck\,2018} : \quad n_{s} = 0.9658 \pm 0.0038 , \quad r < 0.072 \,,\nonumber\\&&
    {\rm BICEP/Keck2021} : \quad r < 0.036 .\nonumber
\end{eqnarray}
Using the upper limit on the tensor-to-scalar ratio $r < 0.036$ at 95\% confidence, we can first constrain $\lambda$ in Eq.(\ref{finalrm2}) to obtain
\ba
\lambda < 5.48\times 10^{-2} N-1\,.
\ea
For example, taking $N=60\,(50)$, it yields $\lambda < 2.29\,(1.74)$. Using parameters of the base $\Lambda$CDM cosmology reported by Planck 2018 for $P_{\cal R}$ at the scale $k = 0.05$\,Mpc$^{-1}$, we find from Eq.(\ref{PR33m1}) that
\begin{eqnarray}
   \alpha &=& \frac{1}{(\lambda  (\lambda +2)+1)}\Bigg(8.4253\times 10^6 N^2 \nonumber\\&&+ ((1.68506\times 10^7 \lambda +1.68506\times 10^7) \log (\Delta_{*} )+2.09361\times 10^7)N^2 \beta \Bigg)\,,\label{alpha}
\end{eqnarray}
where $\Delta_{*}=H_{*}/\mu$ is the parameter at the time when the perturbation with comoving momentum, $k=k_{*}$ crosses the Hubble radius during inflation. The behaviors of $\alpha$ versus $\lambda$ using a set of parameters can be displayed in Fig.(\ref{figal}). In the top panel, we consider various values of $\beta$, ranging from $10^{-4}$ to $10^{-2}$, while keeping $N$ and $\Delta_{*}$ fixed at $N=60$ and $\Delta_{*}=10^{5}$. The $R+R^{2}$ model of inflation in gravity's rainbow was also specified. We find that for $\beta\leq 10^{-3}$, the predictions do not differ from those obtained in the $R+R^{2}$ model, while for $\beta > 10^{-3}$, the predictions deviate from those obtained in the $R+R^{2}$ model \cite{Chatrabhuti:2015mws}. We also obtain the same behaviors in the bottom panel while keeping $\Delta_{*}$ fixed at $\Delta_{*}=10^{7}$.
We observe that when employing $\beta=[10^{-4}, 10^{-3}, 10^{-2}]$ while keeping $N=60$ constant, the coefficient $\alpha$ exhibited magnitudes of $\alpha\sim {\cal O}(10^{10})$ for $\lambda < 1.0$. More specifically, the magnitude of $\alpha$ can be higher by decreasing $\lambda$ displayed in Fig.(\ref{figal}).

\begin{figure}[ht!]
    \centering
\includegraphics[width=5in,height=5in,keepaspectratio=true]{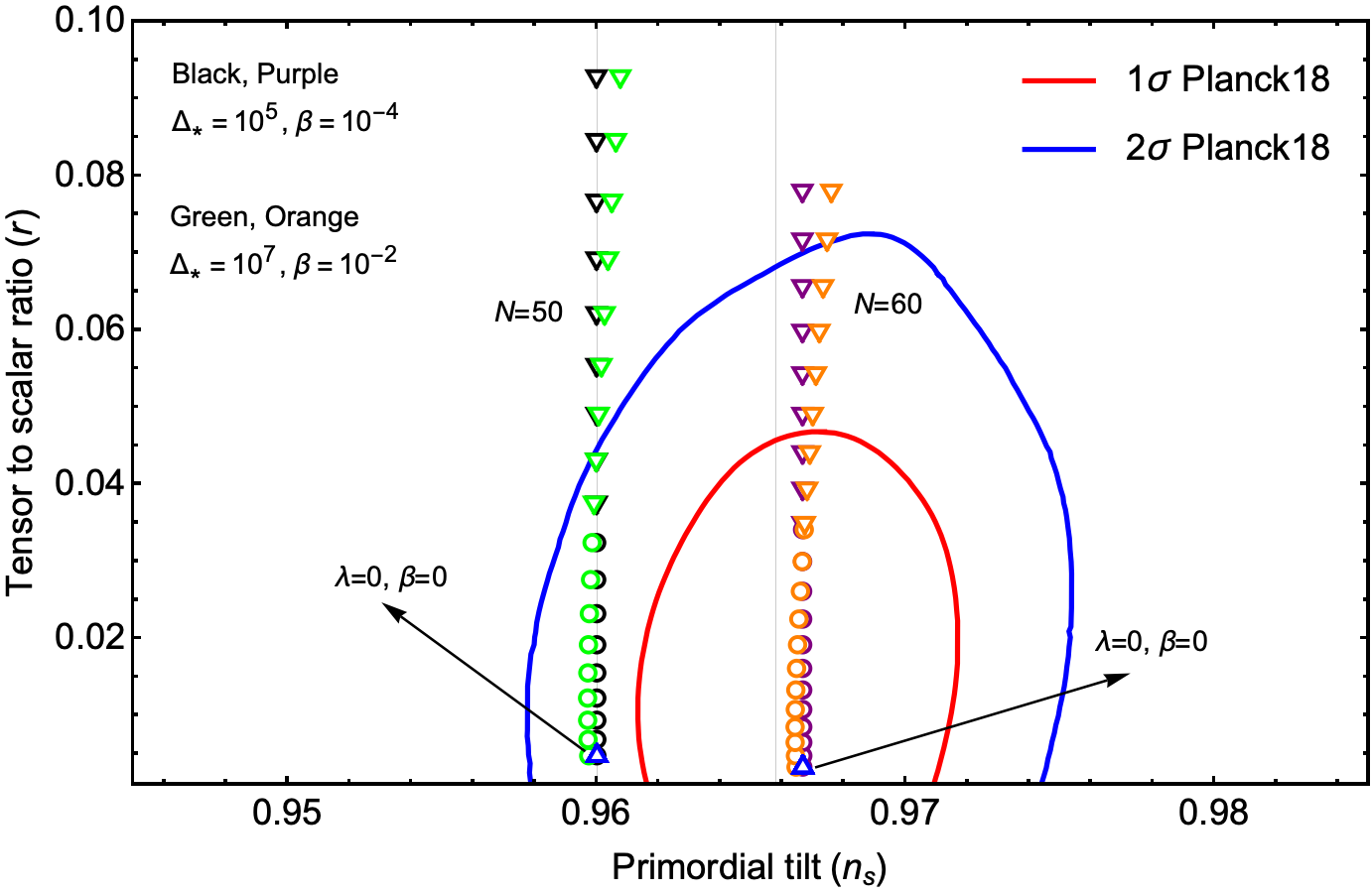}
    \caption{We display the tensor-to-scalar ratio $r$ as a function of the scalar spectral index $n_s$. We consider two different sets of parameters: Left plots: $\Delta_{*}=10^{5},\,\beta=10^{-4}$ (Black) and $\Delta_{*}=10^{7},\,\beta=10^{-2}$ (Green) using $N=50$; Right plots: $\Delta_{*}=10^{5},\,\beta=10^{-4}$ (Purple) and $\Delta_{*}=10^{7},\,\beta=10^{-2}$ (Orange) using $N=60$. The predictions of $R+R^{2}$ has been also identified with $\lambda=0,\,\beta=0$ (Blue). The contours show the allowed values of $n_s$ up to $2\sigma$ confident level.}
    \label{rnss}
\end{figure}

\begin{figure}[ht!]
    \centering
\includegraphics[width=5in,height=5in,keepaspectratio=true]{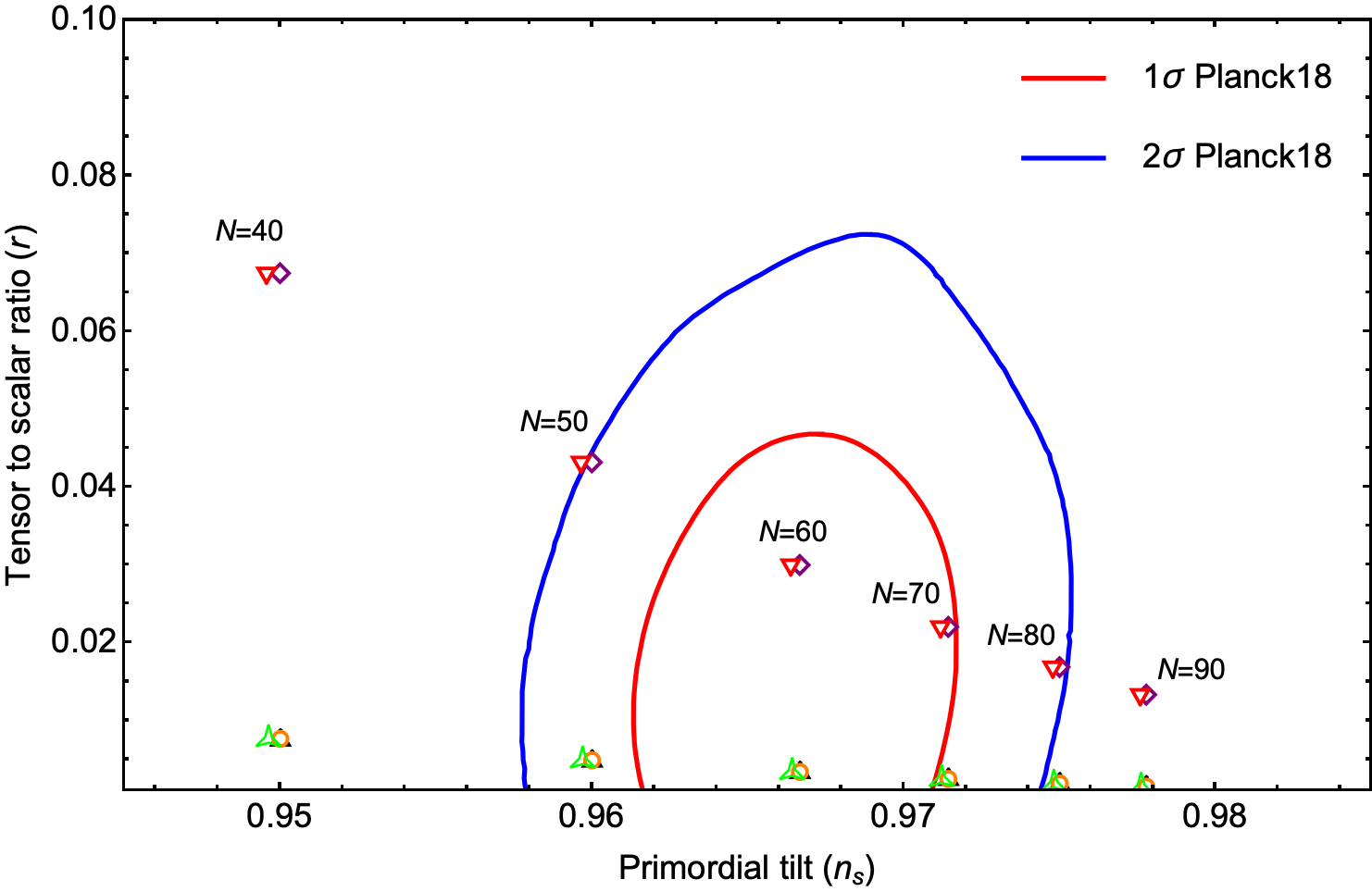}
    \caption{We display the tensor-to-scalar ratio $r$ as a function of the scalar spectral index $n_s$. The Uptriangles represent predictions for $\lambda $=0,\,$\beta $=0. The upper plots show predictions for $\lambda =2.0,\, \Delta_{*}=10^5,\,\beta =10^{-4}$ (Diamonds) and $\lambda =2.0,\, \Delta_{*}=10^5,\,\beta =10^{-2}$ (Downtriangles), while the lower plots show predictions for $\lambda =10^{-2},\, \Delta_{*}=10^5,\,\beta =10^{-4}$ (Circles) and $\lambda =10^{-2},\, \Delta_{*}=10^5,\,\beta =10^{-2}$ (Polygons). We show for numerous values of $N$ from Left to Right. The contours show the allowed values of $n_s$ up to $2\sigma$ confident level. The COBE constraint are imposed and $H_{*}=10^{5}\,\mu$.}
    \label{rnss1}
\end{figure}

In Fig.(\ref{rnss}), we display the tensor-to-scalar ratio $r$ as a function of the scalar spectral index $n_s$. We consider two different sets of parameters: Left plots: $\Delta_{*}=10^{5},\,\beta=10^{-4}$ (Black) and $\Delta_{*}=10^{7},\,\beta=10^{-2}$ (Green) using $N=50$; Right plots: $\Delta_{*}=10^{5},\,\beta=10^{-4}$ (Purple) and $\Delta_{*}=10^{7},\,\beta=10^{-2}$ (Orange) using $N=60$. The predictions of $R+R^{2}$ has been also identified with $\lambda=0,\,\beta=0$ (Blue). The contours show the allowed values of $n_s$ up to $2\sigma$ confident level. Interestingly, the greater the product between $\beta$ and $\Delta_{*}$, the more noticeable the difference from the original model of $R+R^{2}$ in gravity's rainbow \cite{Chatrabhuti:2015mws}.

In Fig.(\ref{rnss1}), we display the tensor-to-scalar ratio $r$ as a function of the scalar spectral index $n_s$. The Uptriangles represent predictions for $\lambda $=0,\,$\beta $=0. The upper plots show predictions for $\lambda =2.0,\, \Delta_{*}=10^5,\,\beta =10^{-4}$ (Diamonds) and $\lambda =2.0,\, \Delta_{*}=10^5,\,\beta =10^{-2}$ (Downtriangles), while the lower plots show predictions for $\lambda =10^{-2},\, \Delta_{*}=10^5,\,\beta =10^{-4}$ (Circles) and $\lambda =10^{-2},\, \Delta_{*}=10^5,\,\beta =10^{-2}$ (Polygons). We show for numerous values of $N$ from Left to Right. The contours show the allowed values of $n_s$ up to $2\sigma$ confident level. The COBE constraint are imposed and $H_{*}=10^{5}\,\mu$. Remarkably, when imposing consistency with observations on the spectral index $n_s$, we ascertain that the tensor-to-scalar ratio $r$ can adhere to the upper limit set by the Planck collaboration, $r<0.036$. This value holds promise for potential measurement by Stage IV CMB ground experiments and is certainly within reach of future dedicated space missions such as LiteBIRD and COrE. Concretely, we find that $r=0.03$ and $n_{s}=0.966$ for $N=60,\,\beta=10^{-4},\,\Delta_{*}=10^{5}$ and $H_{*}=10^{5}\,\mu$.

%%%%%%%%%%%%%%%%%%%%%%%%%%%%%%%%
\section{Concluding Remarks}
%%%%%%%%%%%%%%%%%%%%%%%%%%%%%%%%
In this work, we have considered an effective gravity model (\ref{loa}) motivated by asymptotic safety. This model has been proposed by Refs.\cite{Liu:2018hno,Koshelev:2022olc} as a refined Starobinsky model in the context of gravity as an asymptotically safe theory. Our model can be viewed as an extension of Starobinsky's $R^2$ inflation. In the present work, we have considered the underlying effective Lagrangian within the framework of gravity's rainbow. The COBE normalization and the Planck constraint on the scalar spectrum have been used to constrain the predictions. 

We have demonstrated that the power spectrum of curvature perturbation relies on $\alpha$ and $\beta$, as well as on a rainbow parameter. Similarly, the scalar spectral index $n_s$ is influenced by $\beta$ and the rainbow parameter, yet remains unaffected by $\alpha$. Additionally, the tensor-to-scalar ratio $r$ solely depends on the rainbow parameter. Remarkably, when requiring $n_s$ to be consistent with the Planck collaboration at $1\sigma$ confidence level, the upper limit on the tensor-to-scalar ratio $r < 0.036$ can be naturally satisfied. This value potentially holds promise for potential measurement by Stage IV CMB ground experiments and is certainly within reach of future dedicated space missions such as LiteBIRD and COrE.

%%%%%%%%%%%%%%%%%%%%%%%%%%%%%%%%%%%%%%%%
%%%%%%%%%%%%%%%%%%%%%%%%%%%%%%%%%%%%%%%%

\begin{thebibliography}{99}

%\cite{Demmel:2015oqa}
\bibitem{Demmel:2015oqa}
M.~Demmel, F.~Saueressig and O.~Zanusso,
%``A proper fixed functional for four-dimensional Quantum Einstein Gravity,''
JHEP \textbf{08}, 113 (2015)
%doi:10.1007/JHEP08(2015)113
%[arXiv:1504.07656 [hep-th]].
%121 citations counted in INSPIRE as of 23 Apr 2024

%\cite{Liu:2018hno}
\bibitem{Liu:2018hno}
L.~H.~Liu, T.~Prokopec and A.~A.~Starobinsky,
%``Inflation in an effective gravitational model and asymptotic safety,''
Phys. Rev. D \textbf{98}, no.4, 043505 (2018)
%doi:10.1103/PhysRevD.98.043505
%[arXiv:1806.05407 [gr-qc]].
%49 citations counted in INSPIRE as of 23 Apr 2024

%\cite{Koshelev:2022olc}
\bibitem{Koshelev:2022olc}
A.~S.~Koshelev, K.~S.~Kumar and A.~A.~Starobinsky,
%``Generalized non-local R$^{2}$-like inflation,''
JHEP \textbf{07}, 146 (2023)
%doi:10.1007/JHEP07(2023)146
%[arXiv:2209.02515 [hep-th]].
%15 citations counted in INSPIRE as of 23 Apr 2024

\bibitem{Cherenkov} 
Cherenkov Telescope Array. (2020). Retrieved 13 April 2020, from https://www.cta-observatory.org/

%\cite{Magueijo:2002xx}
\bibitem{Magueijo:2002xx} 
  J.~Magueijo and L.~Smolin,
  %``Gravity's rainbow,''
  Class.\ Quant.\ Grav.\  {\bf 21}, 1725 (2004)
  
    %\cite{Ling:2006az}
\bibitem{Ling:2006az} 
  Y.~Ling,
  %``Rainbow universe,''
  JCAP {\bf 0708}, 017 (2007)
  
  %\cite{Feng:2017gms}
\bibitem{Feng:2017gms} 
  Z.~W.~Feng and S.~Z.~Yang,
  %``Thermodynamic phase transition of a black hole in rainbow gravity,''
  Phys.\ Lett.\ B {\bf 772}, 737 (2017)
  
  %\cite{Hendi:2018sbe}
\bibitem{Hendi:2018sbe} 
  S.~H.~Hendi and M.~Momennia,
  %``AdS charged black holes in Einstein–Yang–Mills gravity's rainbow: Thermal stability and $P − V$ criticality,''
  Phys.\ Lett.\ B {\bf 777}, 222 (2018)
  
  %\cite{Panahiyan:2018fpb}
\bibitem{Panahiyan:2018fpb} 
  S.~Panahiyan, S.~H.~Hendi and N.~Riazi,
  %``$AdS_{4}$ dyonic black holes in gravity's rainbow,''
  Nucl.\ Phys.\ B {\bf 938}, 388 (2019)
  
  %\cite{Dehghani:2018qvn}
\bibitem{Dehghani:2018qvn}
  M.~Dehghani,
  %``Thermodynamics of charged dilatonic BTZ black holes in rainbow gravity,''
  Phys.\ Lett.\ B {\bf 777} (2018) 351
  
  %\cite{Upadhyay:2018vfu}
\bibitem{Upadhyay:2018vfu} 
  S.~Upadhyay, S.~H.~Hendi, S.~Panahiyan and B.~Eslam Panah,
  %``Thermal fluctuations of charged black holes in gravity’s rainbow,''
  PTEP {\bf 2018}, no. 9, 093E01 (2018)
  
  %\cite{Dehghani:2018svw}
\bibitem{Dehghani:2018svw} 
  M.~Dehghani,
  %``Thermodynamics of novel charged dilaton black holes in gravity's rainbow,''
  Phys.\ Lett.\ B {\bf 785}, 274 (2018)
  
  %8
%\cite{Hendi:2017pld}
\bibitem{Hendi:2017pld} 
  S.~H.~Hendi, A.~Dehghani and M.~Faizal,
  %``Black hole thermodynamics in Lovelock gravity's rainbow with (A)dS asymptote,''
  Nucl.\ Phys.\ B {\bf 914}, 117 (2017)

%9
%\cite{Hendi:2016njy}
\bibitem{Hendi:2016njy} 
  S.~H.~Hendi, S.~Panahiyan, B.~Eslam Panah, M.~Faizal and M.~Momennia,
  %``Critical behavior of charged black holes in Gauss-Bonnet gravity’s rainbow,''
  Phys.\ Rev.\ D {\bf 94}, no. 2, 024028 (2016)

%10
%\cite{Hendi:2015cra}
\bibitem{Hendi:2015cra} 
  S.~H.~Hendi, M.~Faizal, B.~E.~Panah and S.~Panahiyan,
  %``Charged dilatonic black holes in gravity’s rainbow,''
  Eur.\ Phys.\ J.\ C {\bf 76}, no. 5, 296 (2016)

%11
%\cite{Hendi:2015hja}
\bibitem{Hendi:2015hja} 
  S.~H.~Hendi and M.~Faizal,
  %``Black holes in Gauss-Bonnet gravity’s rainbow,''
  Phys.\ Rev.\ D {\bf 92}, no. 4, 044027 (2015)


%12
%\cite{Ali:2015iba}
\bibitem{Ali:2015iba} 
  A.~F.~Ali, M.~Faizal, B.~Majumder and R.~Mistry,
  %``Gravitational Collapse in Gravity's Rainbow,''
  Int.\ J.\ Geom.\ Meth.\ Mod.\ Phys.\  {\bf 12}, no. 09, 1550085 (2015)

%13
%\cite{Ali:2014zea}
\bibitem{Ali:2014zea} 
  A.~F.~Ali, M.~Faizal and M.~M.~Khalil,
  %``Remnant for all Black Objects due to Gravity's Rainbow,''
  Nucl.\ Phys.\ B {\bf 894}, 341 (2015)

%14
%\cite{Ali:2014qra}
\bibitem{Ali:2014qra} 
  A.~F.~Ali, M.~Faizal and M.~M.~Khalil,
  %``Absence of Black Holes at LHC due to Gravity's Rainbow,''
  Phys.\ Lett.\ B {\bf 743}, 295 (2015)

%15
%\cite{Ali:2014yea}
\bibitem{Ali:2014yea} 
  A.~F.~Ali, M.~Faizal and M.~M.~Khalil,
  %``Remnants of black rings from gravity`s rainbow,''
  JHEP {\bf 1412}, 159 (2014)

%16
%\cite{Ali:2014cpa}
\bibitem{Ali:2014cpa} 
  A.~F.~Ali, M.~Faizal and B.~Majumder,
  %``Absence of an Effective Horizon for Black Holes in Gravity's Rainbow,''
  EPL {\bf 109}, no. 2, 20001 (2015)

%21
%\cite{EslamPanah:2018ums}
\bibitem{EslamPanah:2018ums} 
  B.~Eslam Panah,
  %``Effects of energy dependent spacetime on geometrical thermodynamics and heat engine of black holes: gravity's rainbow,''
  Phys.\ Lett.\ B {\bf 787}, 45 (2018)
  
  %%%%%BH%%%%%%%
%\cite{Feng:2016zsj}
\bibitem{Feng:2016zsj} 
  Z.~W.~Feng, S.~Z.~Yang, H.~L.~Li and X.~T.~Zu,
  %``Thermodynamics and Phase transition of Schwarzschild black hole in Gravity's Rainbow,''
  arXiv:1608.06824 [physics.gen-ph].
  %%CITATION = ARXIV:1608.06824;%%
  %4 citations counted in INSPIRE as of 03 Mar 2019
  
  %\cite{Hendi:2016hbe}
\bibitem{Hendi:2016hbe} 
  S.~H.~Hendi, S.~Panahiyan, S.~Upadhyay and B.~Eslam Panah,
  %``Charged BTZ black holes in the context of massive gravity’s rainbow,''
  Phys.\ Rev.\ D {\bf 95}, no. 8, 084036 (2017)
  
    %%%%%%Other Application%%%%%%%%
  
  %\cite{Momeni:2017cvl}
\bibitem{Momeni:2017cvl} 
  D.~Momeni, S.~Upadhyay, Y.~Myrzakulov and R.~Myrzakulov,
  %``Cosmic string in gravity’s rainbow,''
  Astrophys.\ Space Sci.\  {\bf 362}, no. 9, 148 (2017)
  
  %\cite{Deng:2017umx}
\bibitem{Deng:2017umx} 
  X.~M.~Deng and Y.~Xie,
  %``Gravitational time advancement under gravity's rainbow,''
  Phys.\ Lett.\ B {\bf 772}, 152 (2017)
  
  %H
%\cite{Khodadi:2016aop}
\bibitem{Khodadi:2016aop} 
  M.~Khodadi, K.~Nozari and B.~Vakili,
  %``Classical and quantum dynamics of a perfect fluid scalar-energy dependent metric cosmology,''
  Gen.\ Rel.\ Grav.\  {\bf 48}, no. 5, 64 (2016)

%\cite{Khodadi:2016bcx}
\bibitem{Khodadi:2016bcx} 
  M.~Khodadi, K.~Nozari and H.~R.~Sepangi,
  %``More on the initial singularity problem in gravity’s rainbow cosmology,''
  Gen.\ Rel.\ Grav.\  {\bf 48}, no. 12, 166 (2016)
  
    %17
%\cite{Rudra:2016alu}
\bibitem{Rudra:2016alu} 
  P.~Rudra, M.~Faizal and A.~F.~Ali,
  %``Vaidya Spacetime for Galileon Gravity's Rainbow,''
  Nucl.\ Phys.\ B {\bf 909}, 725 (2016)


%18
%\cite{Ashour:2016cay}
\bibitem{Ashour:2016cay} 
  A.~Ashour, M.~Faizal, A.~F.~Ali and F.~Hammad,
  %``Branes in Gravity’s Rainbow,''
  Eur.\ Phys.\ J.\ C {\bf 76}, no. 5, 264 (2016)

%19
%\cite{Garattini:2012ec}
\bibitem{Garattini:2012ec} 
  R.~Garattini,
  %``Distorting General Relativity: Gravity's Rainbow and f(R) theories at work,''
  JCAP {\bf 1306}, 017 (2013)

%20
%\cite{Garattini:2014rwa}
\bibitem{Garattini:2014rwa} 
  R.~Garattini and E.~N.~Saridakis,
  %``Gravity’s Rainbow: a bridge towards Hořava–Lifshitz gravity,''
  Eur.\ Phys.\ J.\ C {\bf 75}, no. 7, 343 (2015)
  
  %%%%%%%%%%%%fR gravity%%%%%%%%%%%%%%

  %\cite{Hendi:2016tiy}
\bibitem{Hendi:2016tiy} 
  S.~H.~Hendi, M.~Momennia, B.~Eslam Panah and M.~Faizal,
  %``Nonsingular Universes In Gauss–bonnet Gravity’s Rainbow,''
  Astrophys.\ J.\  {\bf 827}, no. 2, 153 (2016)
  
   %\cite{Hendi:2017vgo}
\bibitem{Hendi:2017vgo} 
  S.~H.~Hendi, M.~Momennia, B.~Eslam Panah and S.~Panahiyan,
  %``Nonsingular universe in massive gravity's rainbow,''
Phys. Dark Universe 16 (2017) 26
  
  %7
%\cite{Heydarzade:2017rpb}
\bibitem{Heydarzade:2017rpb} 
  Y.~Heydarzade, P.~Rudra, F.~Darabi, A.~F.~Ali and M.~Faizal,
  %``Vaidya spacetime in massive gravity's rainbow,''
  Phys.\ Lett.\ B {\bf 774}, 46 (2017)


   %\cite{Hendi:2016oxk}
\bibitem{Hendi:2016oxk} 
  S.~H.~Hendi, B.~Eslam Panah, S.~Panahiyan and M.~Momennia,
  %``F(R) gravity's rainbow and its Einstein counterpart,''
  Adv.\ High Energy Phys.\  {\bf 2016}, 9813582 (2016)
  
    %\cite{Chatrabhuti:2015mws}
\bibitem{Chatrabhuti:2015mws} 
  A.~Chatrabhuti, V.~Yingcharoenrat and P.~Channuie,
  %``Starobinsky Model in Rainbow Gravity,''
  Phys.\ Rev.\ D {\bf 93}, no. 4, 043515 (2016)
  
      %\cite{Codello:2014sua}
\bibitem{Codello:2014sua} 
  A.~Codello, J.~Joergensen, F.~Sannino and O.~Svendsen,
  %``Marginally Deformed Starobinsky Gravity,''
  JHEP {\bf 1502}, 050 (2015)
  
  %\cite{Channuie:2019kus}
\bibitem{Channuie:2019kus}
P.~Channuie,
%``Deformed Starobinsky model in gravity’s rainbow,''
Eur. Phys. J. C \textbf{79} (2019) no.6, 508
%doi:10.1140/epjc/s10052-019-7031-x
%[arXiv:1903.05996 [gr-qc]].
%4 citations counted in INSPIRE as of 18 Apr 2020

%\cite{Waeming:2020rir}
\bibitem{Waeming:2020rir}
A.~Waeming and P.~Channuie,
%``Inflation from f(R) theories in gravity\textquoteright{}s rainbow,''
Eur. Phys. J. C \textbf{80}, no.9, 802 (2020)
%doi:10.1140/epjc/s10052-020-8387-7
%[arXiv:2005.08310 [gr-qc]].
%7 citations counted in INSPIRE as of 17 Apr 2024

%%%%%%%%%%%AS
%\cite{Ferrara:2013rsa}
\bibitem{Ferrara:2013rsa}
S.~Ferrara, R.~Kallosh, A.~Linde and M.~Porrati,
%``Minimal Supergravity Models of Inflation,''
Phys. Rev. D \textbf{88}, no.8, 085038 (2013)
%doi:10.1103/PhysRevD.88.085038
%[arXiv:1307.7696 [hep-th]].
%397 citations counted in INSPIRE as of 23 Apr 2024

%\cite{Asaka:2015vza}
\bibitem{Asaka:2015vza}
T.~Asaka, S.~Iso, H.~Kawai, K.~Kohri, T.~Noumi and T.~Terada,
%``Reinterpretation of the Starobinsky model,''
PTEP \textbf{2016}, no.12, 123E01 (2016)
%doi:10.1093/ptep/ptw161
%[arXiv:1507.04344 [hep-th]].
%27 citations counted in INSPIRE as of 23 Apr 2024

%\cite{Huang:2013hsb}
\bibitem{Huang:2013hsb}
Q.~G.~Huang,
%``A polynomial f(R) inflation model,''
JCAP \textbf{02}, 035 (2014)
%doi:10.1088/1475-7516/2014/02/035
%[arXiv:1309.3514 [hep-th]].
%102 citations counted in INSPIRE as of 23 Apr 2024

%\cite{Ivanov:2021chn}
\bibitem{Ivanov:2021chn}
V.~R.~Ivanov, S.~V.~Ketov, E.~O.~Pozdeeva and S.~Y.~Vernov,
%``Analytic extensions of Starobinsky model of inflation,''
JCAP \textbf{03}, no.03, 058 (2022)
%doi:10.1088/1475-7516/2022/03/058
%[arXiv:2111.09058 [gr-qc]].
%36 citations counted in INSPIRE as of 23 Apr 2024

%\cite{Motohashi:2014tra}
\bibitem{Motohashi:2014tra}
H.~Motohashi,
%``Consistency relation for $R^p$ inflation,''
Phys. Rev. D \textbf{91}, 064016 (2015)
%doi:10.1103/PhysRevD.91.064016
%[arXiv:1411.2972 [astro-ph.CO]].
%60 citations counted in INSPIRE as of 23 Apr 2024

%\cite{Bamba:2015uma}
\bibitem{Bamba:2015uma}
K.~Bamba and S.~D.~Odintsov,
%``Inflationary cosmology in modified gravity theories,''
Symmetry \textbf{7}, no.1, 220-240 (2015)
%doi:10.3390/sym7010220
%[arXiv:1503.00442 [hep-th]].
%298 citations counted in INSPIRE as of 23 Apr 2024

%\cite{Planck:2018jri}
\bibitem{Planck:2018jri}
Y.~Akrami \textit{et al.} [Planck],
%``Planck 2018 results. X. Constraints on inflation,''
Astron. Astrophys. \textbf{641}, A10 (2020)
%doi:10.1051/0004-6361/201833887
%[arXiv:1807.06211 [astro-ph.CO]].
%3001 citations counted in INSPIRE as of 19 Apr 2024

%\cite{CMB-S4:2020lpa}
\bibitem{CMB-S4:2020lpa}
K.~Abazajian \textit{et al.} [CMB-S4],
%``CMB-S4: Forecasting Constraints on Primordial Gravitational Waves,''
Astrophys. J. \textbf{926}, no.1, 54 (2022)
%doi:10.3847/1538-4357/ac1596
%[arXiv:2008.12619 [astro-ph.CO]].
%195 citations counted in INSPIRE as of 23 Apr 2024

%%%%%%%FR
%\cite{Nojiri:2010wj}
\bibitem{Nojiri:2010wj}
S.~Nojiri and S.~D.~Odintsov,
%``Unified cosmic history in modified gravity: from F(R) theory to Lorentz non-invariant models,''
Phys. Rept. \textbf{505}, 59-144 (2011)
%doi:10.1016/j.physrep.2011.04.001
%[arXiv:1011.0544 [gr-qc]].
%3558 citations counted in INSPIRE as of 23 Apr 2024

%\cite{Nojiri:2017ncd}
\bibitem{Nojiri:2017ncd}
S.~Nojiri, S.~D.~Odintsov and V.~K.~Oikonomou,
%``Modified Gravity Theories on a Nutshell: Inflation, Bounce and Late-time Evolution,''
Phys. Rept. \textbf{692}, 1-104 (2017)
%doi:10.1016/j.physrep.2017.06.001
%[arXiv:1705.11098 [gr-qc]].
%1945 citations counted in INSPIRE as of 23 Apr 2024

%\cite{Odintsov:2023weg}
\bibitem{Odintsov:2023weg}
S.~D.~Odintsov, V.~K.~Oikonomou, I.~Giannakoudi, F.~P.~Fronimos and E.~C.~Lymperiadou,
%``Recent Advances in Inflation,''
Symmetry \textbf{15}, no.9, 1701 (2023)
%doi:10.3390/sym15091701
%[arXiv:2307.16308 [gr-qc]].
%85 citations counted in INSPIRE as of 03 May 2024

%\cite{Sotiriou:2008rp}
\bibitem{Sotiriou:2008rp}
T.~P.~Sotiriou and V.~Faraoni,
%``f(R) Theories Of Gravity,''
Rev. Mod. Phys. \textbf{82}, 451-497 (2010)
%doi:10.1103/RevModPhys.82.451
%[arXiv:0805.1726 [gr-qc]].
%3828 citations counted in INSPIRE as of 23 Apr 2024

%\cite{DeFelice:2010aj}
\bibitem{DeFelice:2010aj}
A.~De Felice and S.~Tsujikawa,
%``f(R) theories,''
Living Rev. Rel. \textbf{13}, 3 (2010)
%doi:10.12942/lrr-2010-3
%[arXiv:1002.4928 [gr-qc]].
%3247 citations counted in INSPIRE as of 23 Apr 2024

%\cite{Planck:2018vyg}
\bibitem{Planck:2018vyg}
N.~Aghanim \textit{et al.} [Planck],
%``Planck 2018 results. VI. Cosmological parameters,''
Astron. Astrophys. \textbf{641}, A6 (2020)
[erratum: Astron. Astrophys. \textbf{652}, C4 (2021)]
%doi:10.1051/0004-6361/201833910
%[arXiv:1807.06209 [astro-ph.CO]].
%13432 citations counted in INSPIRE as of 19 Apr 2024

%\cite{BICEP:2021xfz}
\bibitem{BICEP:2021xfz}
P.~A.~R.~Ade \textit{et al.} [BICEP and Keck],
%``Improved Constraints on Primordial Gravitational Waves using Planck, WMAP, and BICEP/Keck Observations through the 2018 Observing Season,''
Phys. Rev. Lett. \textbf{127}, no.15, 151301 (2021)
%doi:10.1103/PhysRevLett.127.151301
%[arXiv:2110.00483 [astro-ph.CO]].
%635 citations counted in INSPIRE as of 19 Apr 2024

\end{thebibliography}
\end{document}